\documentclass[amsmath, amsonfts,twocolumn]{revtex4}

\usepackage[colorlinks]{hyperref}

\pdfoutput=1

\def\minus{%
  \setbox0=\hbox{-}%
  \vcenter{%
    \hrule width\wd0 height \the\fontdimen8\textfont3
  }%
}

\usepackage{graphicx}
\usepackage{amssymb}
\usepackage{wasysym}
\usepackage{stmaryrd}
 
\usepackage{lineno,hyperref}
\usepackage{amsmath,bm}

\renewcommand{\sl}{{\cal S}}

\newcommand{\Bn}{B_n}
\newcommand{\Bt}{B_t}
\newcommand{\Bz}{B_z}

\newcommand{\grad}{{\bm{\nabla}}}

\renewcommand{\c}{n_x}
\newcommand{\s}{n_y}
\newcommand{\bk}{{\bm k}}
\newcommand{\kx}{{\bm k}\cdot\bx}

\newcommand{\dr}[2]{\frac{\partial #1}{\partial #2}}
\newcommand{\drd}[2]{\frac{\partial^2 #1}{\partial #2 ^2}}
\newcommand{\drt}[2]{\frac{\partial^3 #1}{\partial #2^3 }}
\newcommand{\drq}[2]{\frac{\partial^4 #1}{\partial #2^4 }}

\newcommand{\al}{\alpha_\text{\tiny L}}
\newcommand{\at}{\alpha_\text{\tiny T}}

\newcommand{\kt}{k_\text{\tiny T}}

\newcommand{\ct}{c_\text{\tiny T}}
\newcommand{\be}{k}

\newcommand{\Cep}{C_\varepsilon}

\newcommand{\ep}{\varepsilon}

\newcommand{\X}{\Omega}

\newcommand{\Y}{\beta}

\newcommand{\haut}{h_\text{\tiny b}}
\newcommand{\per}{a}

\newcommand{\rhop}{\rho_\text{\tiny b}}
\newcommand{\Ep}{E_\text{\tiny b}}
\newcommand{\nup}{\nu_\text{\tiny b}}

\newcommand{\Es}{E_\text{\tiny S}}
\newcommand{\nus}{\nu_\text{\tiny S}}

\newcommand{\lams}{\lambda_\text{\tiny S}}
\newcommand{\mus}{\mu_\text{\tiny S}}
\newcommand{\rhos}{\rho_\text{\tiny S}}

\newcommand{\Kl}{K} 

\newcommand{\K}{f} 
 
\newcommand{\KL}{f_\text{\tiny C}} 
\newcommand{\KT}{\K_\text{\tiny F}} 

\newcommand{\kah}{\kappa \haut}
\newcommand{\omz}{\omega_0}

\newcommand{\ph}{\varphi}
\newcommand{\ps}{{\bm \psi}}
\newcommand{\bu}{{\bm u}}
\newcommand{\bx}{{\bm x}}

\newcommand{\bn}{{\bm n}}
\newcommand{\bt}{{\bm t}}

\newcommand{\ex}{{\bm e}_x}
\newcommand{\ey}{{\bm e}_y}
\newcommand{\ez}{{\bm e}_z}

\newcommand{\dsp}{\displaystyle}
\def\beq{\begin{equation}}
\def\eeq{\end{equation}}

\renewcommand{\u}{ U}

\newcommand{\rp}{r_\text{\tiny b}}

\newcommand{\toutin}{\left\{\begin{array}{l}}
\newcommand{\toutind}{\left\{\begin{array}{ll}}
\newcommand{\toutint}{\left\{\begin{array}{lll}}
\newcommand{\toutout}{\end{array}\right.}

\begin{document}

\title{
Surface waves from flexural and compressional resonances of beams
}
\author{Jean-Jacques Marigo }
\address{Lab. de M\'ecanique des Solides, Ecole Polytechnique, Route de Saclay, 91120 Palaiseau, France
}
\author{Kim Pham }
\address{
IMSIA, CNRS, EDF, CEA, ENSTA Paris, Institut Polytechnique de Paris,
 828 Bd des Mar\'echaux, 91732 Palaiseau, France
}
\author{Agn\`es Maurel}
\address{Institut Langevin, ESPCI ParisTech, CNRS, 1 rue Jussieu, 75005 Paris, France
}
\author{S\'ebastien Guenneau}
\address{
UMI 2004 Abraham de Moivre-CNRS, Imperial College, London SW7 2AZ, UK
}

\begin{abstract} 
We present a three-dimensional model describing the propagation of  elastic  waves in a soil substrate supporting an array of cylindrical beams  experiencing flexural and compressional resonances. 
The resulting  surface waves are of two types. In the sagittal plane, hybridized Rayleigh waves can propagate except within  bandgaps resulting from a complex  interplay between flexural and  compressional resonances.  We  exhibit a wave decoupled from the hybridized Rayleigh wave which  is the elastic analogue of  electromagnetic spoof plasmon polaritons. This wave with displacements perpendicular to the sagittal plane is sensitive only to  flexural resonances. Similar, yet quantitatively different, physics is demonstrated in a two-dimensional setting involving resonances of plates.   
  \end{abstract}
 
\maketitle

Surface elastic waves can propagate in a soil substrate supporting  a periodic array of resonating  elements. This has been chiefly demonstrated  in the GHz regime with  resonant pillars of typically 1/10 mi\-cro\-meter scale  \cite{khelif2010,achaoui2011,oudich2012,achaoui2013}. 
Considering  meter length scale the  frequency range falls in the spectrum  of seismic waves and in this context, an array of beams on a soil substrate is the canonic idealized configuration used in seismology to illustrate the problem of "site-city interaction"  \cite{gueguen}. 
 From a theo\-re\-ti\-cal point of view, most of the
 models encapsulate the behavior of the resonators  with a single or multi-degree of freedom system,
 resulting  in effective boundary conditions of
 the Robin type for the soil on its own \cite{garova,maznev1,boubou3}.    On the basis of these models new devices of seismic metasurfaces have been shown to efficiently  shield  Rayleigh
 \cite{brule,krodel,colombi,colombi2,colquitt,palermo0} and Love  \cite{palermo1,nous1,nous3} waves.  
 In most cases, only the compressional resonances of the resonators were con\-si\-de\-red. In a recent work, the case  of flexural resonances of beams has been considered \cite{wootton}. However, the study  does not address the configuration of beams in perfect contact with the soil and merely considers motions in the sagittal plane due to  flexural  resonances.  
  \vspace{.1cm}
 
   \begin{figure}[h!]
\centering
\includegraphics[width=.9\columnwidth]{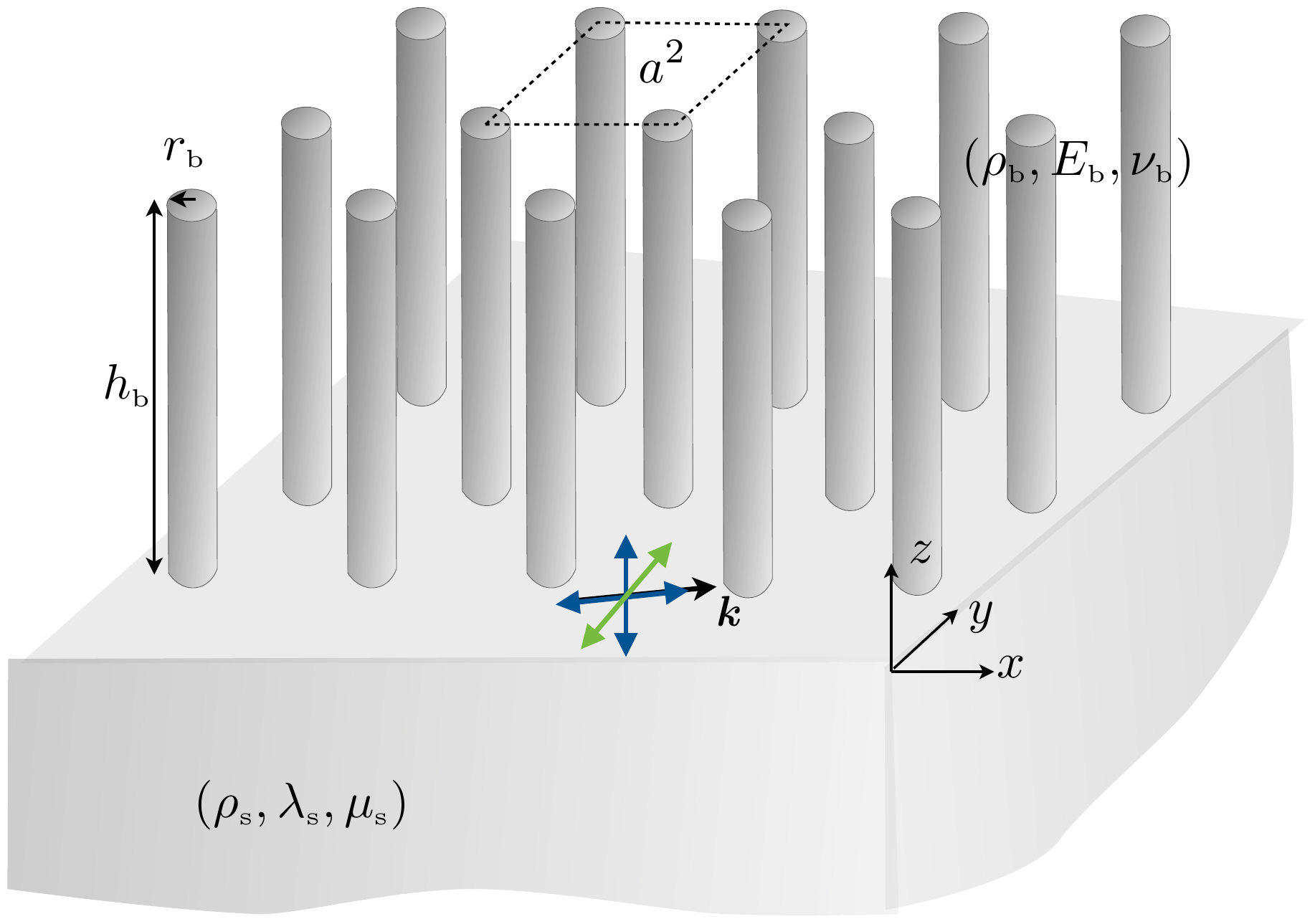}
\caption{ Soil substrate supporting an array of beams with flexural and compressional resonances. Hybridized Rayleigh waves with motions in the sagittal plane (blue arrows show the displacements in the sagittal plan) and out-of-plane elastic SPPs (green arrows).   }
	\label{Fig1}
\end{figure}

 In this Letter we consider this realistic configuration and we propose  a model able to account for both fle\-xu\-ral and compressional resonances in three dimensions, see Figure \ref{Fig1}.   In the sagittal plane,  hybridized Rayleigh waves are found whose dispersion results from a complex interplay between both types of resonances; in particular two Rayleigh waves can coexist at the same frequency.  In addition, a surface wave with displacements perpendicular to the sagittal plane  appears to be  the elastic analog of spoof plasmon polaritons (SPPs) in electromagnetism  \cite{pendry} with a dispersion   governed by the flexural resonances only. 
 The two-dimensional case of an array of parallel plates can be modelled almost identically and unveils  important quantitative differences with the three-dimensional case.

\vspace{.2cm} 
We denote $(\rhop,\Ep,\nup)$ the mass density, Young's mo\-du\-lus and the Poisson ratio of the beams; $(\rhos,\lams,\mus)$ are the mass density and the two Lam\'e's coefficients of the elastic soil substrate. 
When slender beams are considered, meaning that their radius $\rp$ is much smaller than their height $\haut$, a reduction of model from three dimensions to one (vertical) dimension is possible and this results  in the well-known equations  
   \beq
   \toutin
\dsp   \drq{\u_\alpha}{z}-\kappa^4\u_\alpha=0,\quad \alpha=x,y,\\[10pt]
  \dsp \drd{\u_z}{z}+\Kl^2 \u_z=0,\\
   \toutout
\label{tt}   \eeq
where $\u_\alpha$, $\alpha=x,y$, are the horizontal displacements in the region of the beams and $\u_z$ the vertical displacement, see {\em e.g.} \cite{colquitt,wootton}. In \eqref{tt}, the wavenumbers   $\kappa=\left(\frac{4\rhop \omega^2}{\Ep\rp^2}\right)^{1/4}$  and $\Kl=\omega \sqrt{\frac{\rhop}{\Ep}}$ are  associated with flexural and compressional  resonances respectively.
   These equations have to be supplied with boundary conditions at the top of the beams and at their junction with the soil substrate. In the actual problem  stress-free boundary condition apply at the top of the beams and the continuity of the displacements and of the normal stresses applies at the  beam/soil interface. In the reduced model, this leads to  clamped-free effective boundary conditions  namely   \beq\label{bc}\toutind
  \dsp\text{at } z=0: &
  \dsp \u_\alpha=u_\alpha, \quad \dr{\u_\alpha}{z}=0,\quad \u_z=u_z,\\[8pt]
   & \dsp \sigma_{z\alpha}=-\theta \Ep\frac{\rp^2}{4} \drt{\u_\alpha}{z},\quad
   \dsp \sigma_{zz}=\theta \Ep \dr{\u_z}{z},
  \\[6pt]
  \text{at } z=\haut &  \dsp \drd{\u_\alpha}{z}=\drt{\u_\alpha}{z}=0, \quad \dr{\u_z}{z}=0,
   \toutout \eeq 
where $(u_\alpha,u_z)$ are the displacements in the soil, $(\sigma_{z\alpha},\sigma_{zz})$ the associated normal stress and $\theta$  the cross-sectional area ratio of beam or plate in the unit cell of horizontal extent $\per$, see Table \ref{table1}. 
It is worth noting that such conditions can be either postulated as in \cite{wootton} or derived using asymptotic analysis combined with homogenization \cite{nous1,nous3,nous2}.   
\vspace{.2cm}

The conditions of prescribed displacements and zero rotation at $z=0$ together with the conditions of free rotation and free horizontal displacements at $z=\haut$  make it possible to set the problem in the beams as two decoupled linear problems on $\u_\alpha$ and $\u_z$  with respect to  ${u_\alpha}_{|z=0}$ and ${u_z}_{|z=0}$ respectively. 
Denoting $\omz$ the characteristic flexural frequency  and $\ep$ the coupling parameter  
     \beq\label{imp}
\omz=\frac{1}{\sl\haut}\sqrt{\frac{\Ep}{\rhop}},\quad 
\ep=\frac{\theta}{\sl} \sqrt{\frac{\rhop \Ep}{\rhos\mus}},
\eeq 
where $\theta$ and $\sl$ are given in table \ref{table1} (beams), we introduce  the dimensionless frequency $\X=\omega/\omz=(\kah)^2$.
\begin{table}[h!]
\begin{tabular}{| l | c | c  |  }
  \hline
  & slenderness $\sl$ & filling fraction $\theta$  \\[4pt]
  \hline
   3D (beams) & $\sl=2\haut/\rp$ & $\theta=\pi\rp^2/\per^2$ \\
  \hline
   2D (plates) & $\sl=\sqrt{3(1-\nup^2)}\haut/\rp$ & $\theta=2\rp/\per$\\
  \hline
 \end{tabular}
 \caption{Slenderness $\sl$ and filling fraction $\theta$ for beams and plates (with $\per$ the array spacing) entering in \eqref{imp} resulting in the same modelling \eqref{robin}. }
\label{table1} \end{table}

From \eqref{tt}-\eqref{bc},  boundary conditions of the Robin's type are found  for the substrate on its own
\beq\toutin
\dsp     \sigma_{x\alpha}(\bx,0)=\mus\kt \ep\, \KT(\X)\; u_\alpha(\bx,0), \quad \alpha=x,y  \\[10pt]
\dsp   \sigma_{zz}(\bx,0)=\mus \kt \ep  \,\KL(\X) \; u_z(\bx,0),
  \toutout\label{robin} \eeq
 where  $\kt=\omega/\ct$ ($\ct=\sqrt{\mus/\rhos}$) 
and
   \beq
\begin{array}{l}
\dsp    \KT(\X)=\sqrt{\X}\;\frac{\sin\sqrt{\X}\cosh\sqrt{\X}+\cos\sqrt{\X}\sinh\sqrt{\X}}{1+\cosh\sqrt{\X}\cos\sqrt{\X}},\\[10pt]
\dsp    \KL(\X)=\sl \tan\left(\X/\sl \right),
\end{array}  \label{func} \eeq
being  the impedance functions encapsulating the flexural and compressional  resonances of the beams. 

\vspace{.2cm}
  We are looking for a surface wave   evanescent for $z\to-\infty$ and propagating along the interface $z=0$ with a wavevector  $\bk=k\bn$  ($\bn=\c\ex+\s\ey$,  $(\ex,\ey)$ being the unit vectors along $x$ and $y$, and $\c^2+\s^2=1$). The solution is written in terms of  the elastic potentials $(\ph,\ps)$, such that $\bu=\Re\left[\grad\ph+\grad\times \ps\right]$, with $\grad\cdot \ps=0$. Making use of  the isotropy of the medium, we define 
\beq\toutin\label{homo}
\dsp \ph(\bx,z)=-\frac{iA}{k} \,e^{\be \al z +i \kx},\\[10pt] 
\dsp \ps(\bx,z)=\frac{1}{k}
\left(\Bn\bn+\Bt\bt+i\Bz\ez\right) \,e^{\be\at z+i \kx},
\toutout\eeq
 with $\Bn+\at \Bz=0$, and where $\bt=-\s\ex+\c\ey$. With $\Y$ the ratio of the celerities of the Rayleigh wave $c$ and of the bulk shear wave $\ct$, we have 
\beq
\Y=\frac{\kt}{\be}, \quad 1-\at^2=\Y^2, \quad 1-\al^2=\xi\Y^2,
\eeq
 where $\xi=\mus/(\lams+2\mus)$. 
 The resulting expressions of  $(u_\alpha,u_z)$,  $(\sigma_{z\alpha},\sigma_{zz})$ along with  \eqref{robin} provide two decoupled systems,  on the displacements  $(u_n,u_z)$  (with $u_n=\c u_x+\s u_y$  and $\sigma_n=\c \sigma_{xz}+\s \sigma_{yz}$ the associated stress) and  on the out-of-plane displacement $u_t=-\s u_x+\c u_y$ and  $\sigma_t=-\s \sigma_{xz}+\c \sigma_{yz}$ the associated stress.
%
%
The displacements $(u_n,u_z)$ in the sagittal plane correspond  to hybridized  Rayleigh waves whose  dispersion relation reads 
\beq\label{zizi}
\toutin
\left(1+\at^2\right)^2-4\at\al+\Cep(\Y,\X)=0,\\[14pt]
\begin{array}{ll} \Cep(\Y,\X)=& \dsp \ep \Y^3\,\left[
\KL \al+\KT\at\right]+\ep^2\Y^2\KL\KT\left(
\al\at-1\right).
\end{array}\toutout
\eeq
 (where $f$ stands for $f(\X)$). The displacement $u_t$ perpendicular to the sagittal plane is associated with a surface wave whose dispersion
 \beq\label{zuzu}
\Y=\frac{1}{\sqrt{1+\ep^2\KT^2(\X)}}, \quad \KT(\X)\geq 0,
\eeq
is the elastic analog of electromagnetic spoof plasmons \cite{pendry}. Interestingly, such a wave has been announced in a similar  setting involving  Love waves in the presence of a guiding layer (see Fig. 11 in \cite{nous1}).
 As one would expect for $\ep=0$, the classical  Rayleigh waves is recovered, see \eqref{zizi}, and the elastic SPP disappears, see \eqref{zuzu} (with $\be=\kt$). Next, neglecting the flexural resonances ($\KT=0$) produces $\Cep=\ep \Y^3\al \KL$ in agreement with \cite{colquitt}. Eventually,  considering the flexu\-ral resonances in a frequency range well below the first longitudinal resonance gives $\KL(\X)\sim \X$ in agreement with \cite{nous2}.  
 
\vspace{.3cm}
  From now on, we 
set the physical parameters  as follows:  $\Es=0.1$ GPa, $\rhos=10^3$ kg.m$^{-3}$ and 
$\Ep=10\Es$, $\rhop=10\rhos$;  $\nus=\nup=0.3$ (hence $\xi\simeq 0.28$). Next, 
$\rp=0.25$ m and $\per=1$ m and we consider  $\haut= 30$, 15 and 6 m. The resulting parameters entering in \eqref{imp} and \eqref{robin} are given in Table \ref{table2} according to Table \ref{table1}. 
\begin{table}[h!]
\begin{tabular}{| l | c | c  |  c|}
  \hline
  & $\haut=$ 30 m& 15 m &  6 m \\[3pt]
  \hline
   3D &$\begin{array}{c}
   \omz=0.04 \text{ rad.s}^{-1}\\ \ep=1.3\, 10^{-2}\\ \sl=240\end{array}$  & 
   $\begin{array}{c}
   \omz=0.17 \text{ rad.s}^{-1}\\ \ep=2.6 \,10^{-2} \\   \sl=120\end{array}$ & 
   $\begin{array}{c}\omz=1.1 \text{ rad.s}^{-1}\\  \ep=6.6\, 10^{-2} \\ \sl=48\end{array}$  \\
  \hline
2D & $\begin{array}{c}
\omz=0.05  \text{ rad.s}^{-1} \\ \ep=3.8\, 10^{-2} \\ \sl=210
\end{array}$  
 &
 $\begin{array}{c}
\omz=0.21 \text{ rad.s}^{-1} \\ \ep=8.1 \,10^{-2} \\ \sl=102\end{array}$ & 
$\begin{array}{c}
\omz=1.33 \text{ rad.s}^{-1} \\ \ep=20.3 \,10^{-2} \\ \sl=40 \end{array}$  \\
  \hline
 \end{tabular}
 \caption{Reference frequency $\omz$ ($\X=\omega/\omz$) and dimensionless coupling parameter $\ep$ and slenderness $\sl$ entering in  \eqref{robin}-\eqref{func} for beams (3D) and plates (2D), see Table \ref{table1}.}
\label{table2} \end{table}
   
\begin{figure}[h!]
\centering
\includegraphics[width=1\columnwidth]{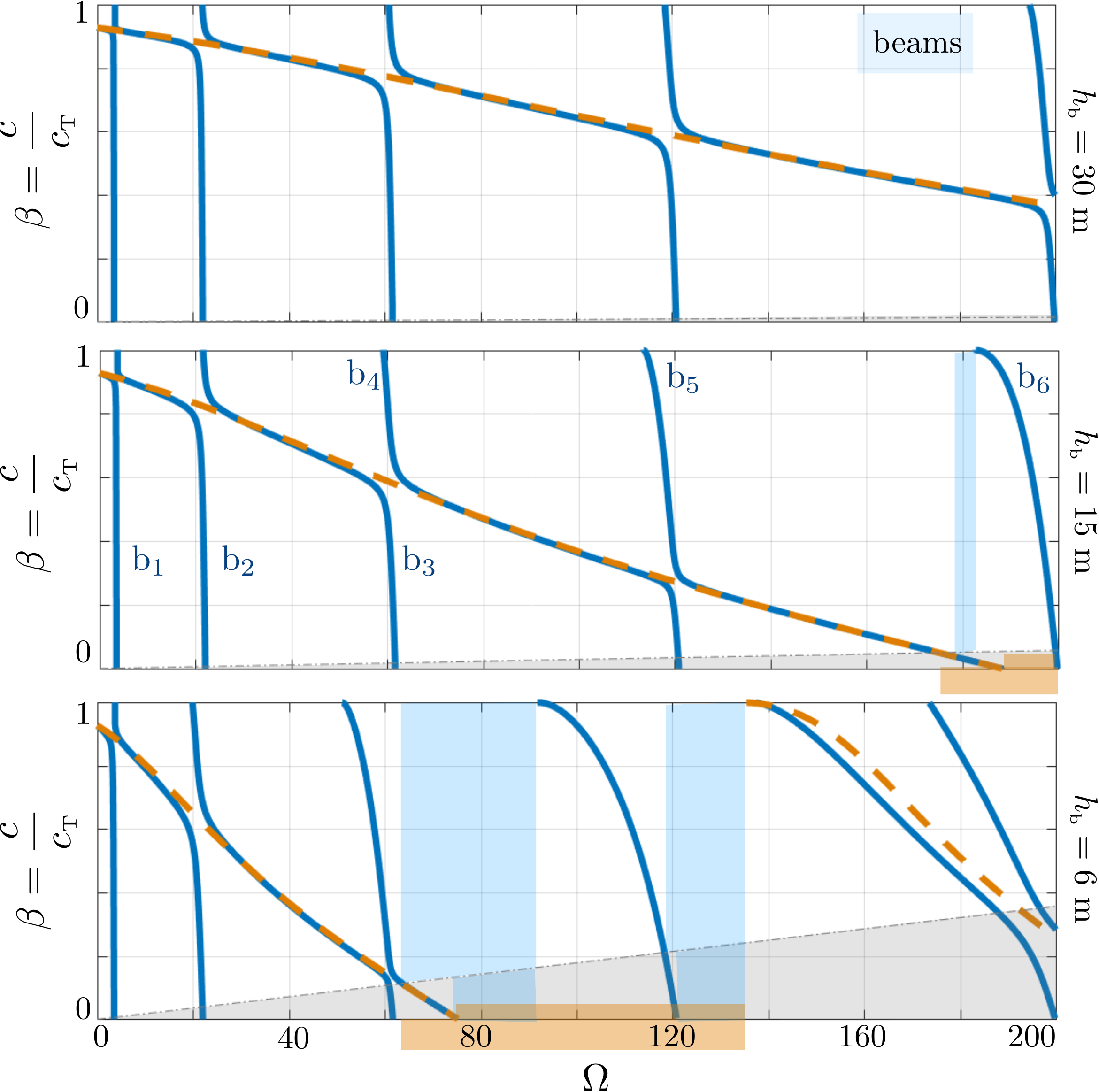}
\caption{ Dispersion  of the hybridized Rayleigh wave for 3D beams  -- dimensionless velocity  $c/\ct$ versus 
$\X=\omega/\omz$ (blue lines). 
 Dashed red lines show the dispersion produced by  the compressional  resonances on their own ($\KT=0$). For $\haut=30$ m, there is  no bandgap and two branches b$_n$ and b$_{n+1}$ coexist in a frequency range just below the $n^\text{\tiny th}$ flexural resonance. For $\haut=15$ m a bandgap is opened by periodicity (light blue region in the Brillouin zone) and at $\haut=6$ m the two  intrinsic bandgaps (light blue region outside of the Brillouin zone,  independent of $\per$) are slightly enlarged by periodicity. }
	\label{Fig2}
\end{figure}

 We report in figure \ref{Fig2} the dispersion relations $\Y(\X)$ of the hybridized Rayleigh waves (blue lines) for decreasing beam heights $\haut$   while keeping constant the range  $\X\in(0,200)$. This allows to keep  the  first 6 flexural resonances at constant values $\X=$ 3.5, 22.0, 61.8, 75.3, 120.9 and 199.8 corresponding to $(1+\cos\sqrt{\X}\cosh\sqrt{\X})=0$. Within this interval,  the dimensionless compressional frequencies $\X_\text{\tiny C}=\frac{(2n+1)\pi}{2}\sl$, $n=0,1,\cdots$,  decrease linearly with  
 $\haut$ while   their frequencies  $\omega_\text{\tiny C}=\sqrt{\frac{\Ep}{\rhop}}\frac{(2n+1)\pi}{2\haut}$ increase (but more slowly than the flexural resonance frequencies).  
 For a large slenderness $\haut=30$ m, the first longitudinal resonance is sent  
to $\X\simeq 377$ ($\omega_\text{\tiny C}=16.5$ Hz) hence $\KL(\X)\simeq \X$ in  \eqref{zizi}.  In this case,  it is easy to see that   the $(n+1)^\text{\tiny th}$ branch , $n=1,\cdots$, of hybridized Rayleigh wave appears (for $\Y=1$) 
before the  $n^\text{\tiny th}$ branch has reached its asymptote (for $\Y=0$). The salient consequence is that the 
 branches $n$ and $(n+1)$ coexist  below the $n^\text{\tiny th}$ flexural resonance frequency. 
 Decreasing  the slenderness with $\haut=15$ m and 6 m leads to the appearance of the first compressional resonance frequency 
at $\X_\text{\tiny C}=188.5$ ($\omega_\text{\tiny C}=33.1$ Hz) and  $\X_\text{\tiny C}=75.4$ ($\omega_\text{\tiny C}=82.8$ Hz). In these cases, relatively small bandgaps (light blue regions ) are opened within the large bandgap dictated by  the compressional resonance on its own (light red regions), revealing the interplay between the two types of resonances. Incidentally, as it is the rule, these intrinsic bandgaps (independent of the array spacing $\per$) can be enlarged by periodicity which imposes $\Y\geq \left(\frac{\omz a}{\pi\ct}\right) \X$ ($k<\pi/\per$) as seen for $\haut=6$ m; also,  bandgaps  solely due to periodicity can be  opened as can be seen for $\haut = 15$ m near $\X=180$.  
\begin{figure}[b!]
\centering
\includegraphics[width=1\columnwidth]{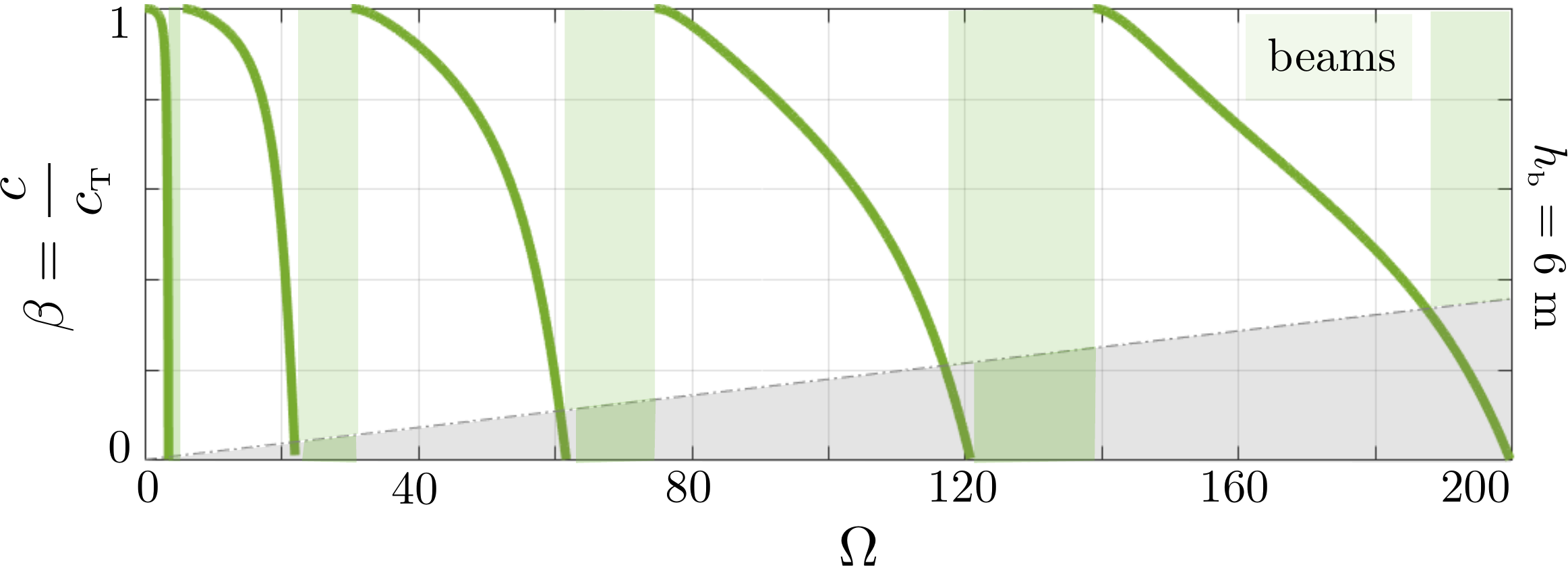}
\caption{ Dispersion  of elastic SPPs, from \eqref{zuzu} with out-of-plane motion for $\haut=6$ m. Intrinsic band-gaps are opened due to flexural resonances (light green regions outside of the Brillouin zone) which are enlarged by periodicity (light green regions inside the Brillouin zone). }
	\label{Fig3}
\end{figure} 
Eventually, in figure \ref{Fig3} we report the dispersion relation of the elastic SPPs, from \eqref{zuzu}, for $\haut=$ 6 m. As their electromagnetic counterparts, these waves have bandgaps  dictated by the condition $\KT\geq 0$ hence the bandgap positions and thicknesses are independent of the coupling $\ep$ and of the slenderness $\sl$. 
As previously said,  similar waves with out-of-plane displacements have  been reported in the presence of a guiding layer  \cite{nous1}, and they are recovered here in the absence of any guiding layer.
\vspace{.2cm}

We now restrict ourselves on to a two-dimensional setting invol\-ving arrays of plates.  As one would expect, the coupling of  plates with the soil is higher than that of  beams for   the same thickness $2\rp$ and the same height $\haut$ (see table \ref{table2}); the slenderness $\sl$ on the contrary is slightly lower which is also expected since plates have a higher flexural rigidity than beams, and this  is translated in an effective lower slenderness, see table \ref{table1}.   Accordingly,  the relative positions of the compressional and flexural resonances change and we shall see that this strongly affects  the dispersion of the hybridized Rayleigh waves. 
We provide in this case a numerical validation of the dispersion curves by means of diverging reflection coefficients computed below the sound line $k<\kt$ using a multimodal 
method \cite{petitmonstre}.
   \begin{figure}[b!]
\centering
\includegraphics[width=1\columnwidth]{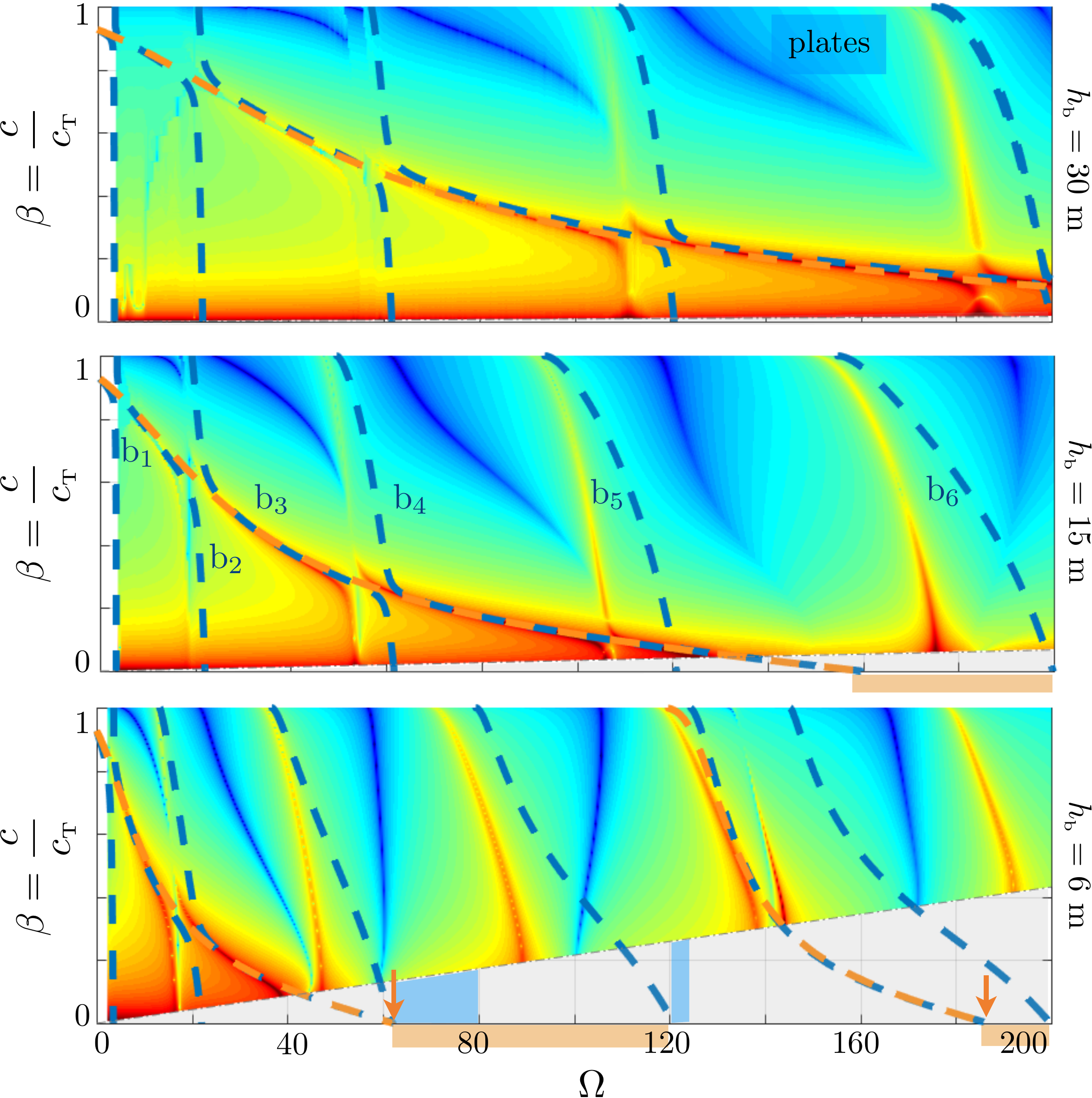}
\caption{  Dispersion  of the hybridized Rayleigh wave for 2D plates  -- The dispersion relations are obtained numerically in the actual problems by means of the divergence of the reflection coefficient (in logarithmic color scale with maxima in red, arbitrary scales have been used). Dashed blue lines show the dispersion from \eqref{zizi}, red vertical arrows for $\haut=15$ m and 6 m indicate the occurence of compressional resonances.  }
	\label{Fig4}
\end{figure}
Results are shown in figure \ref{Fig4}. The overall agreement is good for $\haut=30$ and 15 m although the condition of zero group velocity at the boundary of the Brillouin zone (vertical slope $\frac{\partial \Y}{\partial_\X}=\infty$) produces  more or less pronounced shifts of the branches to lower frequencies \cite{mercier}. For $\haut=6$ m, the agreement is good up to $\X\sim 50$ which can be partially attributed to other resonances and in particular that of an  edge mode for 
$\sqrt{\frac{\rhop}{\Ep}} \omega\haut\simeq 2.32$ ($\X=91$) \cite{pagneux}.  

\begin{figure}[b!]
\centering
\includegraphics[width=1\columnwidth]{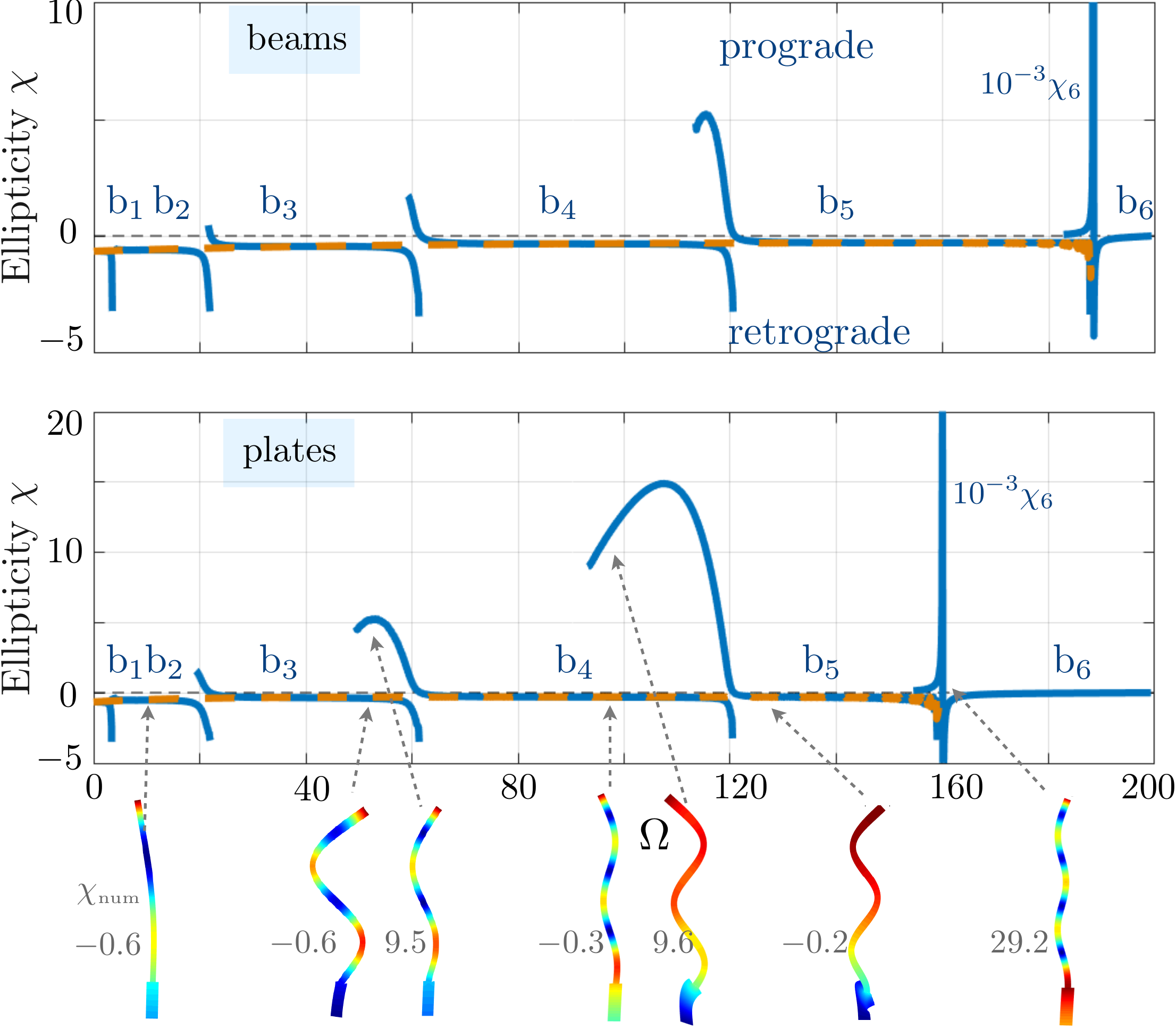}
\caption{  Ellipticity $\chi$ of the hybridized Rayleigh waves versus $\X$ for beams (upper panel) and for plates -- Blue lines show $\chi$ from \eqref{el}, dashed red lines neglecting the flexion ($\KT=0$ in \eqref{el}).  Insets show the deformations in the plates atop the soil in a single cell, computed numerically and the corresponding ellipticity $\chi_\text{\tiny num}$ (see main text).}
	\label{Fig5}
\end{figure}

\vspace{.2cm}
The ellipticity $\chi$ of surface  waves of the 
Rayleigh type (or $H/V$ for horizontal to vertical ratio)  characterizes the displacements at the soil surface, namely  $\chi=a/b$ for displacement at $z=0$ of the form 
$u_n=a\cos(\kx-\omega t)$, $u_z=b\sin(\kx-\omega t)$ and it is an important indicator of the ground motion. From \eqref{zizi}, it reads
\beq\label{el}
\chi=\frac{\at\Y^2+\ep\Y \KL(\al\at-1)}{2\al\at-(1+\at^2)}=\frac{2\al\at-(1+\at^2)}{\Y^2\al+\ep\Y \KT(\al\at-1)},
\eeq
whose variations versus $\X$ are reported  in figure \ref{Fig5} for $\haut=15$ m (in the case of beams and plates). 
In the reported cases, the branches b$_n$, $n=1$ to 5, are below  the first compressional resonance. Making use of \eqref{el} along with \eqref{zizi}, these branches have 3 typical points: (i) the starting point ($\Y=1$) for which $\chi=\ep\KL$ which  increases when $n=1,\cdots, 5$ increases as $\KL$ does; 
 (ii) the frequency at which the wave motion transitions from prograde to retrograde with $\chi=0$ 
when $\KT\to\infty$ and  this is consistent with \eqref{robin} which predicts $u_n=0$ at the flexural resonances, (iii) the ending point for $\Y\to 0$ resulting in $\chi=-1/\xi$ for any branch. Next
the branch $b_6$ appears just before the longitudinal resonance where $\KL\to\infty$ imposes $u_z=0$. It results  high values of $\chi$ in the neighborhood of the singularity and it is worth noting that this affects a branch which is dictated by the interplay of flexural and compressional resonances (this branch is missed if flexural motions are disregarded).  
For comparison, we have computed numerically the displacements at the free surface and in the plates for $\X=13$ (b$_2$), $\X=50$ 
(coexistence of b$_3$ and b$_4$), $\X=94$ (coexistence of b$_4$ and b$_5$), $\X=122$ (b$_5$)
and $\X=156$ (b$_6$).
Results, in the insets of figure \ref{Fig5}, show the deformations in the plates and the corresponding ellipticity  $\chi_\text{\tiny num}$. 
The agreement with \eqref{el} is  qualitative which 
is partly attributable to the existence of boun\-da\-ry layers at the junction between the plates and the soil. However they confirm the main trends of the model, in particular it  is  noticeable  that the motions in the plates are dominantly horizontal even for very low ellipticity.  

\vspace{.2cm}
Beams and plates
atop a soil substrate impact propagation of seismic waves in a richer way than their acoustic counterparts.
 This is due to a complex interplay between the compressional and longitudinal  resonances and to their associated spectra. In particular, (i) the dispersion of hybridized Rayleigh waves shows a important part of the spectrum associated with celerities larger than that of the classical Rayleigh waves;  (ii) These waves are associated with prograde or retrograde motion at the soil interface with large variations of the ellipticity; (iii) The existence of an out-of-plane surface wave, with infinite ellipticity on its own, 
 sheds new light on the analysis of the displacement components  in particular on records of the ambient noise for which the horizontal displacement is the sum of the two contributions \cite{ojo,lott}.

\end{document}